\documentclass[aps,prb,citeautoscript,superscriptaddress,amsmath,amssymb,twocolumn]{revtex4}
\usepackage{float}
\usepackage[caption = false]{subfig}
\usepackage{graphicx}
\usepackage{amsmath}
\usepackage{hyperref}
\usepackage{color}
\usepackage{bigints}
\usepackage{mathrsfs}
\usepackage{algorithm}
\usepackage{algpseudocode}
\usepackage{multirow}
\usepackage{tabularx}
\usepackage{bm}
\definecolor{darkblue}{RGB}{8,81,156}
\hypersetup{colorlinks,breaklinks,
            linkcolor=darkblue,urlcolor=darkblue,
           anchorcolor=darkblue,citecolor=darkblue}

\date{\today}

\allowdisplaybreaks

    \definecolor{dark-purple}{RGB}{118,42,131}
    \definecolor{dark-green}{RGB}{27,120,55}
    \definecolor{light-purple}{RGB}{231,212,232}
    \definecolor{LIGHT-PURPLE}{RGB}{194,165,207}
    \definecolor{light-green}{RGB}{168,216,183}
    \definecolor{gray}{RGB}{186,186,186}
    \definecolor{super-dark-green}{RGB}{0,69,41}
    \definecolor{super-dark-purple}{RGB}{63,0,125}
    \definecolor{super-dark-blue}{RGB}{8,48,107}
    \definecolor{super-dark-red}{RGB}{165,0,38}
    
    \definecolor{super-dark-purple}{RGB}{64,0,75}
    \definecolor{super-dark-green}{RGB}{0,68,27}

\usepackage{tabularx,booktabs} 

\usepackage{array}
\usepackage{longtable}
\newcolumntype{L}[1]{>{\raggedright\let\newline\\\arraybackslash\hspace{0pt}}p{#1}}
\newcolumntype{C}[1]{>{\centering\let\newline\\\arraybackslash\hspace{0pt}}m{#1}}
\newcolumntype{R}[1]{>{\raggedleft\let\newline\\\arraybackslash\hspace{0pt}}m{#1}}

\begin{document}

\title{How to quantify and avoid finite size effects in computational studies of crystal nucleation: The case of homogeneous crystal nucleation}

\author{Sarwar Hussain, Amir Haji-Akbari}
\email{amir.hajiakbaribalou@yale.edu}
\affiliation{Department of Chemical and Environmental Engineering, Yale University, New Haven, CT  06520}

\date{\today}

\begin{abstract}
\noindent
Finite size artifacts arise in molecular simulations of  nucleation when critical nuclei are too close to their periodic images. A rigorous determination of what constitutes too close is, however, a major challenge. Recently, we devised rigorous heuristics for detecting such artifacts based on our investigation of how system size impacts the rate of heterogeneous ice nucleation (Hussain,  Haji-Akbari, \emph{J. Chem. Phys.} \textbf{154}, 014108, \textbf{2021}).
We identified the prevalence of critical nuclei spanning across the periodic boundary, and the thermodynamic and structural properties of the liquid occupying the inter-image region as indicators of finite size artifacts.
Here, we further probe the performance of such heuristics by examining the dependence of homogeneous crystal nucleation rates in the Lennard-Jones liquid on system size. The rates depend non-monotonically on system size and vary by almost six orders of magnitude for the range of system sizes considered here. We confirm that the prevalence of spanning critical nuclei is the primary indicator of finite size artifacts and almost fully explains the observed variations in rate. Proximity, or structuring of the inter-image liquid, however, is not as strong of an indicator due to the fragmented nature of crystalline nuclei.  As a result, the dependence of rate on system size is subtle for the systems with a minuscule fraction of spanning critical nuclei. These observations indicate that our heuristics are universally applicable to different modes of nucleation (homogeneous and heterogeneous) in different systems even if they might be overly stringent for homogeneous nucleation,~e.g.,~in the LJ system. 
\end{abstract}
\maketitle

\section{Introduction}
\label{section:intro}

\noindent
Finite size effects refer to artifacts that arise in molecular simulations due to the finite size of the simulation box, and can lead to deviations from the thermodynamic limit in estimates of thermodynamic~\cite{BinderFerroelectrics1987, MonJChemPhys1992, HorbachPhysRevE1996, AguadoJChemPhys2001, OreaJCP2005, MastnyJChemPhys2007, BiscayJCP2009, BurtJPhysChemC2016}, structural~\cite{SalacusePhysRevE1996} and transport~\cite{YehJPCB2004,  BotanMolPhys2015, JamaliJChemTheoryComput2018} properties as well as rates of rare events such as nucleation.\cite{HussainJCP2021, MichaelidesChemRev2016} Moreover, they can distort the mechanisms of collective phenomena in unphysical ways.\cite{MeadleyJCP2012} There is, however, always a range of system sizes beyond which such effects either fully disappear, or at least become minimal. As such, finite size effects can be simply avoided by simulating really large systems. Resorting to such a conservative approach, however, is only practical for the simplest of systems with inexpensive interatomic potentials.\cite{SwopePhysRevB1990, EnglishPhysRevE2015} Moreover, a system size that one might deem "very large`` based on some physical intuition might still be too small to be devoid  of finite size effects. It is therefore important to devise rigorous metrics or heuristics for detecting and correcting for the existence of finite size effects, and to identify system sizes that are sufficiently large to be free of strong finite size artifacts. It must, however, be emphasized that what constitutes sufficiently large not only depends on the underlying system and the employed force-field, but also on the property or process of interest. For instance, while systems of a few hundred molecules might be large enough for estimating thermodynamic and structural properties of simple liquids~\cite{BinderFerroelectrics1987, MonJChemPhys1992, HorbachPhysRevE1996, SalacusePhysRevE1996, JamaliJChemTheoryComput2018, MandellJStatPhysics1976}, much larger systems might be necessary for probing collective phenomena such as self-assembly,\cite{HajiAkbariEtAl2009} crystal nucleation,\cite{SwopePhysRevB1990, HoneycuttChemPysLett1984, HoneycuttJPC1986, LeyssaleChemPhysLet2003, HussainJCP2021} condensation\cite{PerezJCP2011} and cavitation.\cite{MeadleyJCP2012} As such, rigorously determining the threshold size beyond which strong artifacts are absent is  a  daunting task particularly for the latter category wherein the process of interest can induce large-range correlations that can in turn lead to strong finite size effects, even in systems that appear to be sufficiently large.\cite{HussainJCP2021}

One such collective phenomenon that has been extensively studied using molecular simulations is crystal nucleation. Indeed, efforts to detect and quantify finite size effects in crystal nucleation studies have been ongoing for decades.\cite{MichaelidesChemRev2016, KatarinaJCP2021} The primary source of finite size artifacts in crystal nucleation is the unphysical correlation between a crystalline nucleus and its periodic image(s).  This can either lead to the formation of nuclei that span across the box boundary~\cite{MeadleyJCP2012, StattPRL2015} or can result in unphysical confinement within the metastable phase sandwiched between the nucleus and its periodic images.\cite{MandellJStatPhysics1976, HoneycuttChemPysLett1984, HoneycuttJPC1986, PengJCP2010} The latter effect can be particularly consequential since  crystalline nuclei can induce long-range order within the liquid beyond the crystal surface due to the diffuse and poorly defined nature of a solid-fluid interface. These, alongside other sources of finite size effects, such as solute depletion in mixtures\cite{SalvalaglioPNAS2015, LiuMolPhys2018} and peculiarities of the employed ensembles,\cite{WedekindJCP2006, PerezJCP2011} can significantly alter the kinetics of both homogeneous and heterogeneous nucleation, and have also been found to impact crystal growth.\cite{BurkeJCP1988, OMalleyPRL2003, StreitzPRL2006} Indeed, several earlier computational studies of crystal nucleation~\cite{Matsumoto2002, JungwithJPCB2006, JungwirthJMolLiq2007, JungwirthJPhysChemC2010} are either demonstrated to be~\cite{SanzJACS2013} or are suspected of~\cite{HajiAkbariPNAS2017, HajiAkbariJCP2017} being impacted by strong finite size effects, and many  more recent works have grappled with the issue of justifying the absence of such artifacts in their calculations.\cite{CoxFaraday2013, CoxJCP2015, CoxIIJCP2015, SossoJPhysChemLett2016, LupiJCP2016}

Historically, early efforts~\cite{HoneycuttChemPysLett1984, HoneycuttJPC1986, SwopePhysRevB1990, HuitemaPhysRevB2000} to characterize finite size effects that originate from periodic boundary conditions focused on homogeneous nucleation within the Lennard-Jones (LJ) liquid.\cite{LJProcRSoc1924} Such studies were mostly conducted under deep supercoolings at which nucleation rates are very large and nucleation events are easily accessible to conventional molecular dynamics (MD) simulations. The main conclusion that emerged from these earlier works was that finite size effects can only be confidently avoided for systems that are at least three orders of magnitude larger than the characteristic volume of a critical nucleus. This stringent criterion has, however, been devised under conditions in which a large number of nucleation events can simultaneously occur even within a small simulation box and the ensuing nuclei can join through a spinodal-like mechanism. Its applicability to lower supercoolings at which nucleation events are less frequent is, however, questionable. Because of this, and due to the fact that it is impractical to satisfy it for more complex systems due to prohibitive computational costs, this criterion has been by and large ignored in the literature, and researchers have instead resorted to one of the following two  approaches: (i) testing the robustness to finite size effects by conducting computationally tractable simulations of larger systems,\cite{CoxFaraday2013, CoxJCP2015, CoxIIJCP2015} and (ii) ensuring that the average distance between critical nuclei and  their periodic images does not exceed half the box dimensions.\cite{SossoJPhysChemLett2016, LupiJCP2016} Apart from its larger computational cost, the first approach might lead to inaccurate conclusions as the apparent lack of sensitivity of rate to system size might be coincidental and might not hold if more simulations are conducted over a wider range of system sizes.  The second approach leads to what is generally known as the  "10\% rule`` in the case of homogeneous nucleation, according to which finite size effects will be minimal if the number of particles within the critical nucleus is not larger than 10\% of the total number of molecules/particles within the metastable fluid. This criterion, however, has not been rigorously tested and is not based on anything but qualitative physical intuition. Moreover, the size of the critical nucleus is always fairly sensitive to the utilized classification and clustering algorithm.\cite{NgugenJPhysChemB2015}

In order to address these issues, we recently devised a set of rigorous heuristics to detect and quantify finite-size effects in crystal nucleation simulations.\cite{HussainJCP2021} As our first case study, we focused on heterogeneous ice nucleation within supported nanofilms of monoatomic water (mW)~\cite{MolineroJPCB2009} in the vicinity of model structureless~\cite{MagdaJCP1985} ice nucleating particles (INPs), and identified features of critical nuclei that indicate the extent of finite size effects in a particular rate calculation.  These features, which include the likelihood of spanning across the simulation box, the proximity to periodic images and the properties of the supercooled liquid sandwiched between the nuclei and their periodic images in cases where no spanning occurs, are not system-specific and can be universally applied to crystal nucleation in any system. In this work, we test the performance of these heuristics for homogeneous nucleation in the LJ system. As opposed to heterogeneous nucleation for which the dimensions of the external crystal nucleating surface, $L$, is a proper measure of  system size, in homogeneous nucleation system size is defined as the total number of molecules/particles ($N_p$) in the simulation box. We use our recently developed jumpy forward flux sampling (jFFS)~\cite{HajiAkbariJCP2018} algorithm to calculate homogeneous nucleation rates for ten different systems comprised of $864$ to $55,296$ LJ particles and find the nucleation rate to vary by almost six orders of magnitude. The bulk of this variation, however, is observed when the critical nuclei span across the periodic boundary,~i.e.,~in what we characterized as the spanning regime in Ref.~\citenum{HussainJCP2021}. Beyond this spanning regime, the rates are virtually insensitive to system size, commensurate with other indicators of finite size artifacts such as proximity and inter-image liquid properties. Incidentally, the smallest system for which the nucleation rate does not deviate significantly from that estimated for the largest system follows the \emph{ad hoc} 10\% rule.

This paper is organized as follows. Section~\ref{section:methods} describes the technical details of the conducted molecular dynamics simulations and rate calculations, as well as the methodology utilized for characterizing proximity. We present our results in Section~\ref{section:results} while Section~\ref{section:conclusions} is reserved for concluding remarks.

\section{Methods}
\label{section:methods}

\subsection{System description and molecular dynamics simulations}

\noindent
We consider homogeneous nucleation within the bulk LJ liquid with the interatomic potential truncated at $r_c=2.5\sigma$. All quantities are reported in reduced LJ units. All MD simulations are conducted in the isothermal isobaric ($NpT$) ensemble at $p^*=p\sigma^3/\epsilon=0$ and $T^*=k_BT/\epsilon=0.5$ using the  the Large-scale Atomic/Molecular Massively Parallel Simulator (LAMMPS)~\cite{PimptonLAMMPS1995} package. Equations of motion are  integrated using the velocity-Verlet algorithm with a time step of $\Delta{t}=0.0025\sigma\sqrt{m/\epsilon}$, while temperature and pressure are controlled using the  Nos\'e-Hoover~\cite{NoseMolPhys1984, HooverPhysRevA1985} thermostat and the Parrinelo-Rahman barostat~\cite{ParrinelloJAppPhys1981} with time constants of $10^2\Delta{t}$ and $10^3\Delta{t}$, respectively. 

For homogeneous nucleation, the proper measure of system size is $N_p$, the total number of LJ particles within the simulation box, which varies between $N_p=864$ and $N_p=55,296$ in this work.  In order to generate starting configurations for each system size, we first generate a simulation box with $n_c\times n_c\times n_c$ unit cells of the face-centered cubic (fcc) lattice at a reduced density of $\rho^*=\rho\sigma^3=1.2$ for $n_c=6, 7, 8, 9, 10, 11, 12, 16, 20$ and $24$, and melt each configuration at a reduced temperature of $0.8$ and a reduced pressure of $0.1$ for a total duration of $4\times10^6$ time steps (or $10^4\sigma\sqrt{m/\epsilon}$), saving configurations every $2\times10^4$ time steps. This culminates in 200 configurations, which we then quench down to the target temperature and pressure over $5.5\times10^4$ time steps.  These 200 quenched configurations are then used as starting points for the basin exploration component of jFFS.\cite{HajiAkbariJCP2018} A list of all system sizes and their average box dimensions are listed in Table~\ref{table:rates}.

\begin{table*}[ht]
\newcolumntype{C}{>{\centering\arraybackslash}X}
\caption{Homogeneous crystal nucleation rates computed at  $T^*=0.5$ and $p^*=0$ using molecular dynamics and jumpy forward flux sampling. $N_p$ refers to the number of LJ particles in each system, $\langle L \rangle$ is the average box dimension in the liquid and $N^{*}$ is the critical nucleus size determined using the approach described in Section~\ref{section:jFFS calculations}. The error bars in $\log_{10} \mathcal{J}$ correspond to $95\%$ confidence intervals, while the uncertainties in $N^*$ correspond to the range of $\lambda$'s for which $0.35\le p_c(\lambda)\le 0.65$.} 
\centering
\begin{tabularx}{0.75\textwidth}{C C C C}
\hline\hline
$N_p$ & $\langle L \rangle$ & $N^*$  & $\log_{10}\mathcal{J}$ \\
\hline
864 & 9.93 & $179\pm12$ & $-12.8496\pm0.0781$\\
1,372 & 11.58  & $256\pm15$ & $-15.7278\pm0.0812$\\
2,048 & 13.24 & $321\pm18$ & $-17.5947\pm0.1371$\\
2,916 & 14.89 &  $354\pm22$ & $-18.2812\pm0.0776$\\
4,000 & 16.54 & $376\pm23$ & $-18.8138\pm0.0774$ \\
5,324 & 18.20  &  $376\pm21$ & $-18.8461\pm0.0772$\\	
6,912 & 19.86  &   $380\pm24$ & $-18.9453\pm0.0696$ \\
16,384 & 26.47 & $378\pm23$ & $-18.8687\pm0.0803$\\
32,000 & 33.09 &  $382\pm24$ & $-18.5911\pm0.0774$\\
55,296 & 39.70 &  $380\pm24$ & $-18.6180\pm0.0840$\\
\hline\hline
\end{tabularx}
\label{table:rates}
\vspace{-10pt}
\end{table*}

\subsection{Rate Calculations}

\noindent
We estimate the nucleation rates using forward flux sampling (FFS),\cite{AllenFrenkel2006} which has been extensively used in recent years for studying a wide variety of rare events.\cite{HussainJChemPhys2020} In particular, we use jFFS,\cite{HajiAkbariJCP2018} a generalized variant of FFS that has been specifically optimized for studying aggregation phenomena such as nucleation. In what follows, we discuss the technical details of our jFFS calculations. 

\subsubsection{Order parameter}
\noindent
Most advanced sampling techniques, including jFFS, require an order parameter, $\lambda:\mathcal{Q}\rightarrow\mathbb{R}$, a mathematical function that quantifies the progress of the corresponding rare event for any configuration $x$ within the configuration space $\mathcal{Q}$. Commensurate with the established practice in the field and similar to our earlier works,\cite{HajiAkbariFilmMolinero2014, HajiAkbariPNAS2015, GianettiPCCP2016, HajiAkbariPNAS2017, HussainJACS2021} we choose the number of particles within the largest crystalline nucleus in the system as our order parameter. We first categorize each particle as solid-like or liquid-like based on the classification criterion discussed below. We then cluster the neighboring solid-like particles into crystalline nuclei of different sizes, with the number of particles within the largest cluster chosen as the order parameter.

The classification criterion utilized here was first introduced in Ref.~\citenum{DelagoJCP2008}  and is based on the Steinhardt bond order parameter.\cite{SteinhardtPRB1983}  For each particle $i$, we first compute $\textbf{q}_6(i)\equiv(q_{6,-6}(i),q_{6,-5}(i),\cdots,q_{6,6}(i))$ as,
\begin{eqnarray}
q_{6m} (i) &=& \frac{1}{N_b(i)} \sum_{j=1}^{N_b(i)} Y_{6m}(\theta_{ij},\phi_{ij}),~~-6 \leq m \leq 6
\end{eqnarray}
where $N_b(i)$ is the number of particles that are within a distance $r_{c,n}=1.4\sigma$ from $i$,   $\theta_{ij}$ and $\phi_{ij}$ are the polar and azimuthal angles corresponding to the separation vector $\textbf{r}_{ij}=\textbf{r}_j-\textbf{r}_i$ connecting particle $i$ to its $j^{\text{th}}$ neighbor, and $Y_{6m}(\cdot,\cdot)$'s are the spherical harmonic functions. We then compute $\overline{\textbf{q}}_6(i)$, the neighbor-averaged form of  $\textbf{q}_6(i)$ as,
\begin{eqnarray}
\mathbf{\overline{q}_6}(i) = \frac{1}{1+N_b(i)}\sum_{j=0}^{N_b(i)} \mathbf{q_6}(j)
\end{eqnarray}
where $j=0$ corresponds to the $i^{\text{th}}$ particle itself. The next step is to estimate $\overline{q}_{6}(i)$, which is a scalar invariant of ${\overline{\textbf{q}}_6}(i)$ given by,
\begin{eqnarray}
\overline{q}_6(i) = \sqrt{\frac{4\pi}{13}\sum_{m=-6}^{6}|\overline{q}_{6,m}(i) |^2}
\end{eqnarray}
and classify $i$ as solid-like when $\overline{q}_6(i) \geq 0.3$. Finally, we cluster all the solid-like particles that are within a distance of $r_{c,n}=1.4\sigma$ of one another, and use the size of the largest such cluster as the FFS order parameter. 

\subsubsection{jFFS calculations}
\label{section:jFFS calculations}

\noindent
The order parameter discussed above is jumpy,~i.e.,~that it undergoes high-amplitude fluctuations along short MD trajectories. We therefore compute the nucleation rates using jFFS,\cite{HajiAkbariJCP2018} a generalized variant of FFS that we recently developed for studying processes such as nucleation that can only be described using jumpy order parameters. Similar to conventional FFS, the transition region between the supercooled liquid basin, $A:=\{x\in\mathcal{Q}:\lambda(x)<\lambda_A\}$ and the crystalline basin, $B:=\{x\in\mathcal{Q}:\lambda(x)\ge\lambda_B\}$ is partitioned in jFFS into $N$ regions separated by level sets of $\lambda(\cdot)$ known as milestones. In order to simplify implementation, we follow the algorithm detailed in Sec. III B 1 of Ref.~\citenum{HajiAkbariJCP2018} that involves on-the-fly placement of milestones.

The first stage of jFFS involves sampling $A$ by launching long conventional MD trajectories from each of the 200 starting configuration, and monitoring for first crossings of $\lambda_0$. The The configurations corresponding to these crossings $\mathcal{C}_0=\left\{x_1^{(0)},x_2^{(0)},\cdots,x_{N_0}^{(0)}\right\}$ are stored and $\Phi_0$, the flux of trajectories leaving $A$ and crossing $\lambda_0$ is estimated as
\begin{eqnarray}
\Phi_0 &=& \frac{N_0}{t_b\langle V\rangle},
\end{eqnarray}
Here, $t_b$ is the combined duration of all MD trajectories and always exceeds $2.5\times10^5\sigma\sqrt{m/\epsilon}$ in this study. $\langle V \rangle$, however, is the average volume of the simulation box. 

The next step is to compute $P(\lambda_B|\lambda_0)$, the probability that a trajectory that starts at $\lambda_0$ can reach $\lambda_B$ before returning to $A$. First, the next milestone $\lambda_1$ is chosen so that $\lambda_1>\max_{x\in\mathcal{C}_0}\lambda(x)$, and a total of $M_1$ trajectories are initiated from configurations randomly chosen from within $\mathcal{C}_0$. Each trajectory is terminated upon crossing $\lambda_1$ or returning to $A$, and $P(\lambda_1|\lambda_0)$ is estimated as $N_1/M_1$ where $N_1$ is the number of trial trajectories crossing $\lambda_1$. This procedure is repeated (i.e.,~by choosing the next milestone and estimating the transition probability) until the transition probability converges to unity at which point the system has overcome the nucleation barrier.  The nucleation rate $\mathcal{J}$ is then estimated as
\begin{eqnarray}
\mathcal{J} &=& \Phi_0\prod_{i=1}^N P(\lambda_i|\lambda_{i-1}),
\end{eqnarray}
The statistical uncertainty of the computed $\mathcal{J}$ is estimated using the approach described in Ref.~\citenum{FrenkelFFS_JCP2006}. All reported error bars correspond to 95\% confidence intervals, i.e.,~twice the standard error obtained from this approach. Since the main premise of this work is to resolve subtle changes in rate due to changing the system size, we terminate each iteration only after a minimum of 5,000 crossings of the target milestone.

It has been previously demonstrated in numerous  studies~\cite{LupiJCP2016, LupiNature2017} that the size of the largest crystalline nucleus is a good reaction coordinate for crystal nucleation. This implies that the critical nucleus size can be accurately determined from the committor probability given by
\begin{eqnarray}
p_c(\lambda_i) &=& \prod_{q=i+1}^N P(\lambda_q|\lambda_{q-1}).
\end{eqnarray}
More precisely, $N^*$, the critical nucleus size is obtained by fitting $p_c(\lambda)$ to the following expression:
\begin{eqnarray}
p_c(\lambda) &=& \frac12\Bigg\{1+\text{erf}\left[a(\lambda-N^*)\right]\Bigg\}.
\end{eqnarray}
For all the structural analysis conducted in this work, we consider the configurations collected at jFFS milestones that are the closest to $N^*$ as critical. The committor probabilities for all such configurations are between 50\% and 76\% for the simulations conducted in this work. 

\subsection{Characterizing the proximity of crystalline nuclei to their periodic images}
\label{methods:characterizing proximity}

\noindent
In order to determine whether a crystalline nucleus $\mathcal{N}$ comprised of $m$ particles with positions $\mathcal{N}=\{\textbf{r}_1, \textbf{r}_2, \cdots, \textbf{r}_m\}$  spans across the periodic boundary, we first generate an undirected graph of all particles within $\mathcal{N}$ that are first nearest neighbors  of each other,~i.e.,~whose distance is smaller than $r_{c,n}$. We then label each particle in $\mathcal{N}$ using a set of three integers $\bm{\xi} = (\xi_x, \xi_y, \xi_z)$ that determine the side of the periodic boundary that it resides on with respect to a reference particle. (Note that in our earlier work on heterogeneous nucleation,\cite{HussainJCP2021} only two integers, $\xi_x$ and $\xi_y$, were needed since crystalline nuclei could only cross the periodic boundary along the $x$ and $y$ dimensions.) In the beginning, these labels are all set to $\mathbf{0}=(0, 0, 0)$ except for an arbitrary reference particle $j$ whose label is set to $\mathbf{1}=(1, 1, 1)$. We then initiate a depth-first search across the nearest neighbor graph of all particles in $\mathcal{N}$ starting from $j$. During this search, whenever an unvisited particle $q$ is first visited by following an edge from $p$, we set ${\bm \xi}_q$  to ${\bm \xi}_p$ if $p$ and $q$ are on the same side of the periodic boundary, otherwise the index corresponding to the boundary that is crossed is multiplied by $-1$. For instance, if going from $p$ to $q$ results in a crossing of the $z$ boundary, we set $\bm{\xi}_q:= (\xi_{p,x},\xi_{p,y},-\xi_{p,z})$. The search is stopped and $\mathcal{N}$ is declared as spanning when a pair of previously visited particles are identified that are at the same side of the boundary, but have differing $\bm{\xi}$ indices.  If no such pair is obtained after searching through the entire graph, then the  $\mathcal{N}$ is concluded to be non-spanning.

To quantify the proximity of $\mathcal{N}$ to their periodic images, we first identify its six closest periodic neighbors along the box side vectors $\textbf{b}_x,\textbf{b}_y$ and $\textbf{b}_z$:
\begin{eqnarray}
\mathcal{N}_{n_x n_y n_z}  &=& \{ \mathbf{r}_i + n_x \mathbf{b}_x + n_y \mathbf{b}_y  + n_z \mathbf{b}_z \}_{i=1}^m,
\end{eqnarray}
with $(n_x,n_y,n_z)\in\{(\pm1,0,0), (0,\pm1,0), (0,0,\pm1)\}$. We then identify particles that have identical $\bm\xi$ labels and calculate the smallest intervals wherein the $x, y$ and $z$ coordinates of particles within each group reside. These intervals are subsequently used to shift the entire cluster so that all its particles lie within the same side of the boundary. After shifting $\mathcal{N}$, we compute the inter-image vector $\textbf{u}$ as,
\begin{eqnarray}
[\tilde{i}, \tilde{j}, \tilde{n}_x, \tilde{n}_y, \tilde{n}_z] &=& \text{argmin}_{i,j \leq m, n_x, n_y, n_z} \left|\mathbf{r}_i - \mathbf{r}^{n_x, n_y, n_z}_j\right| \notag \\
&& \\
\mathbf{u} &=& \mathbf{r}^{\tilde{n}_x, \tilde{n}_y, \tilde{n}_z}_{\tilde{j}} - \mathbf{r}_{\tilde{i}}
\end{eqnarray}
The liquid that resides along $\mathbf{u}$ is referred to as the \emph{inter-image liquid}, and its density profile, $\rho_{\mathbf{u}}(r)$ is calculated by enumerating the average number of LJ particles that reside within  cylindrical bins of radius $r_{c,n}$ and reduced thickness  $\sigma/10$ along the inter-image vector $\mathbf{u}$.

\begin{figure}
\centering
\vspace{-5pt}
\includegraphics[width=0.4\textwidth]{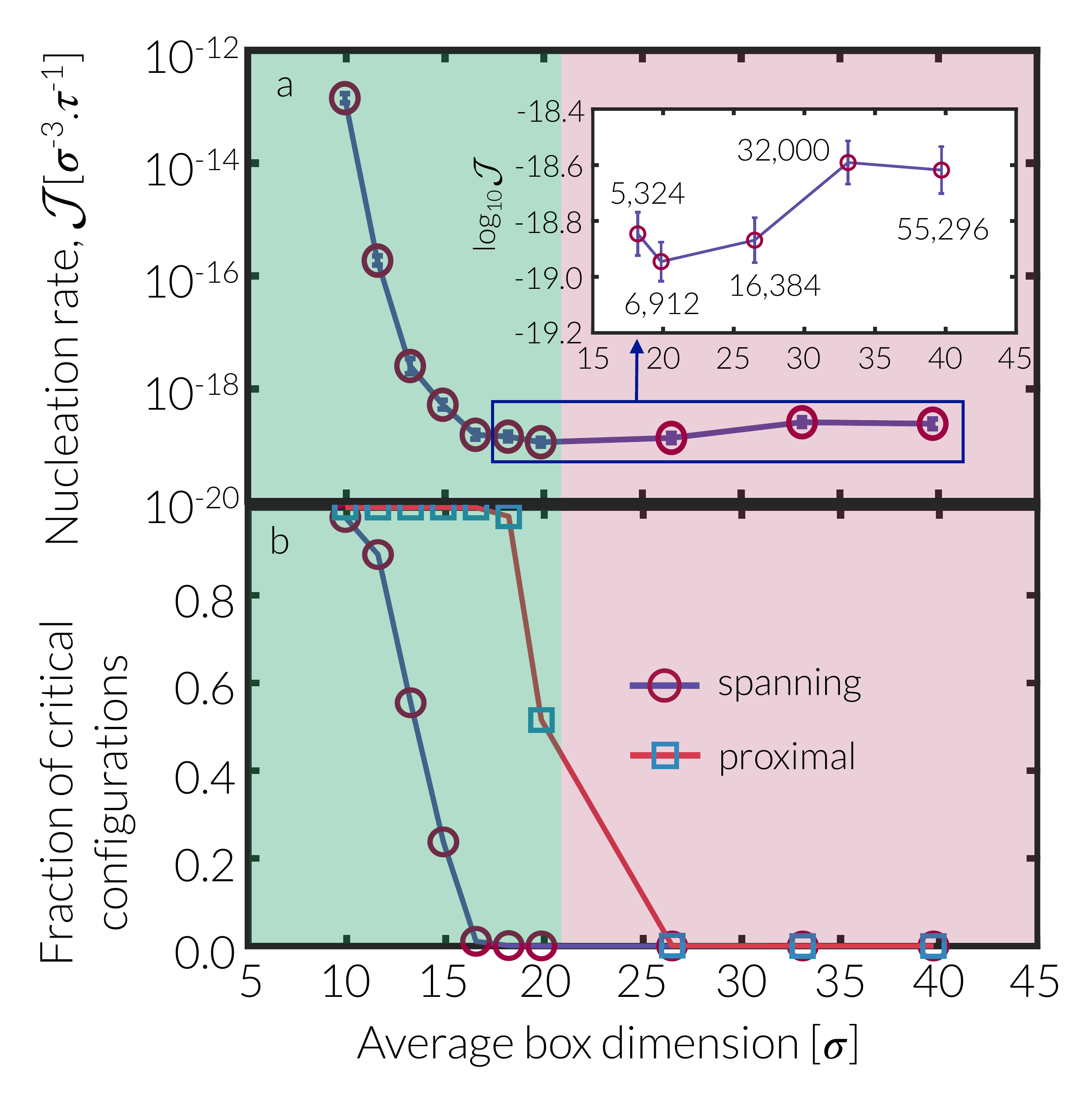}
\vspace{-10pt}
\caption{(a) Homogeneous nucleation rate as a function of $\langle L\rangle$, the average box dimension in the LJ system at $T^*=0.5$ and $p^*=0$.  Labels next to the data points in the inset correspond to the number of particles in each system. (b) The fraction of critical configurations that are spanning (circles) and proximal (squares). Error bars correspond to 95\% confidence intervals and are smaller than the symbols.}
\label{Figure 1}
\vspace{-10pt}
\end{figure}

\subsection{Characterizing cluster compactness}
\label{section:methods_IQ}

\noindent
In order to quantify the compactness of crystalline nuclei, we compute a quantity called \emph{isoperimetric quotient (IQ)}, which has been previously used to estimate the sphericity of polyhedra~\cite{PolyaPrinceton1954, PabloScience2012} and is defined as
\begin{eqnarray}
IQ = 36\pi\frac{V^2}{S^3}.
 \end{eqnarray}
Here, $V$ and $S$ are the volume and the surface area of the object, respectively. IQ values can range from zero to unity, with a value of one obtained for a perfect sphere while zero is only possible for a degenerate planar object with $V=0$. In order to compute IQ, we first construct the Voronoi tessellation of all LJ particles using the \texttt{voro++} library.\cite{RycroftChaos2009} The volume of the cluster is then defined as the sum of volumes of Voronoi cells of its constituent particles, while its surface area is estimated as the sum of the areas of liquid-exposed facets of those Voronoi cells (i.e., facets shared between a nucleus-based Voronoi cell and that of a neighboring liquid-like particle).

\section{Results and Discussions}
\label{section:results}
\subsection{Summary of Nucleation Rates}
\label{results:summary}

\noindent
Fig.~\ref{Figure 1}a depicts the homogeneous nucleation rate, $\mathcal{J}$, as a function of $N_p$ at  $T^*=0.5$ and  $p^*=0$.  (The computed rates are also given in Table~\ref{table:rates}.) The  rates exhibit a non-monotonic dependence on system size, which is qualitatively similar to what we previously observed in the case of heterogeneous ice nucleation~\cite{HussainJCP2021}. We also identify two distinct regimes for the dependence of rate on $N_p$, highlighted with shaded red and green in Fig.~\ref{Figure 1}. For smaller systems, 
the rate is a strong function of system size and changes by as much as six orders of magnitude, indicative of strong finite size effects. For larger systems, however, nucleation rates depend weakly on $N_p$ and only vary by less than a third of an order of magnitude, as shown in the inset of Fig.~\ref{Figure 1}a, suggesting that finite size effects are minimal.  Our analysis of the geometric properties of the critical nuclei reveals that these two regimes differ in the existence of critical nuclei that span across the periodic boundary,~i.e,~that connect with their periodic images to form infinitely sized unphysical nuclei. While no spanning critical nucleus is observed in the red region of Fig.~\ref{Figure 1}, at least one critical nucleus is found to span the periodic boundary in the green region with the fraction of spanning critical nuclei increasing as system size decreases. We therefore refer to the green and red regions of Fig.~\ref{Figure 1} as the \emph{spanning} and \emph{non-spanning} regimes, respectively.  In what follows, we discuss the peculiarities of these two regimes separately.

\subsection{The spanning regime}
\label{results:spanning}

\begin{figure}
\centering
\includegraphics[width=0.5\textwidth]{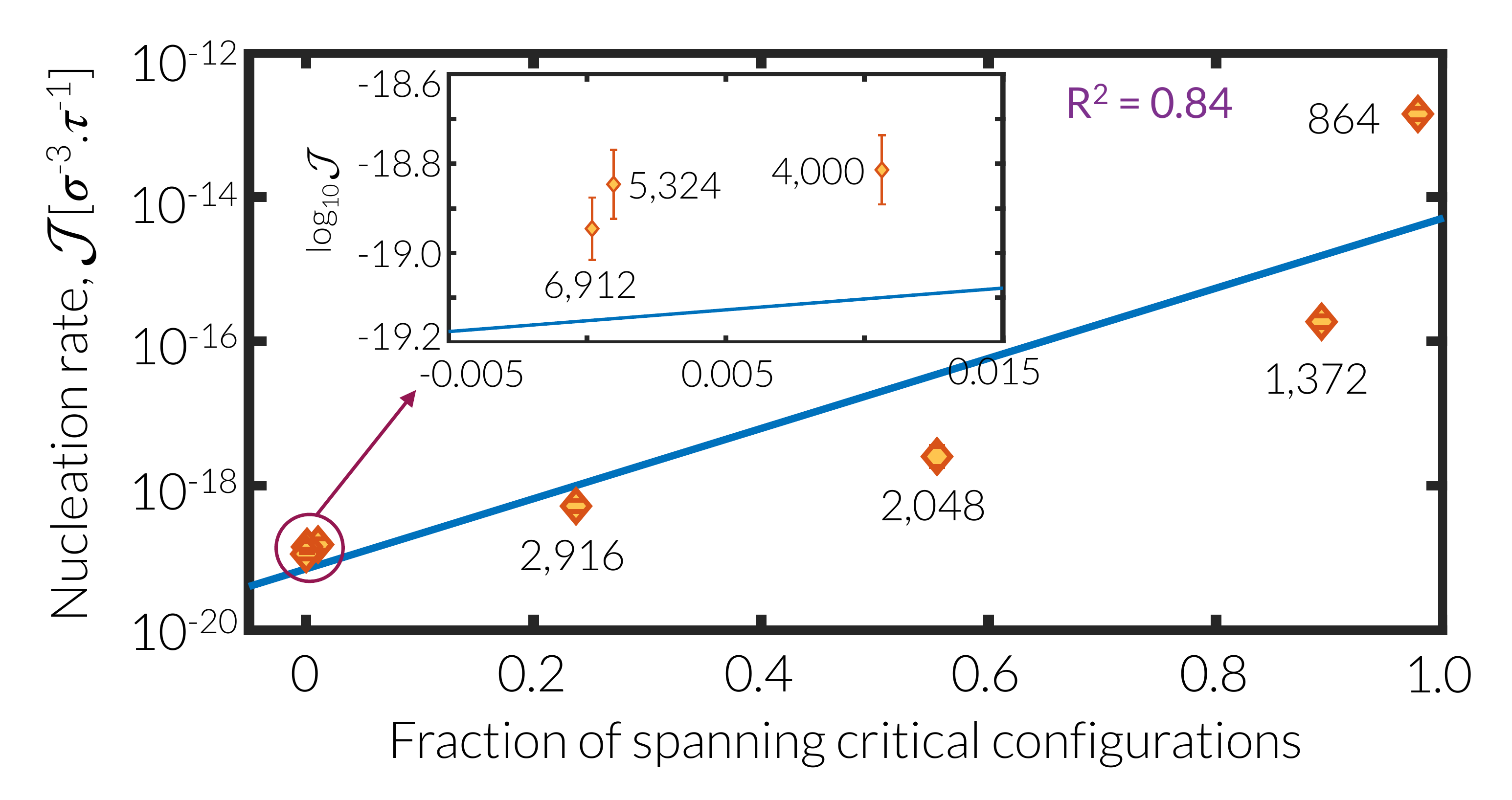}
\vspace{-15pt}
\caption{Linear correlation between the fraction of critical spanning nuclei and $\log_{10}\mathcal{J}$ in the spanning regime. Labels next to data points correspond to the number of particles within each system.}
\vspace{-10pt}
\label{Figure 2}
\end{figure}

\noindent
The spanning regime is a vivid confirmation of our intuition that close proximity between critical nuclei and their periodic images will result in unphysical confinement and a strong dependence of rate on system size, as spanning across the periodic boundary is the most extreme example of such proximity. As can be seen in Fig.~\ref{Figure 1}a and Table~\ref{table:rates}, an increase in the fraction of spanning critical nuclei leads to a considerable increase in nucleation rate and a noticeable decrease in the  size of the critical nucleus. This is because the formation of spanning nuclei results in  an artificial promotion of nucleation. 
Indeed, a strong linear correlation exists between the fraction of spanning critical nuclei and $\log_{10}\mathcal{J}$, as depicted in Fig.~\ref{Figure 2}. This observation is similar to what we previously reported for the spanning regime of  heterogeneous ice nucleation~\cite{HussainJCP2021} and suggests that irrespective of the mode of nucleation, the existence of spanning critical nuclei is a key indicator of potentially strong finite size artifacts.

\begin{figure}
\centering
\includegraphics[width=0.5\textwidth]{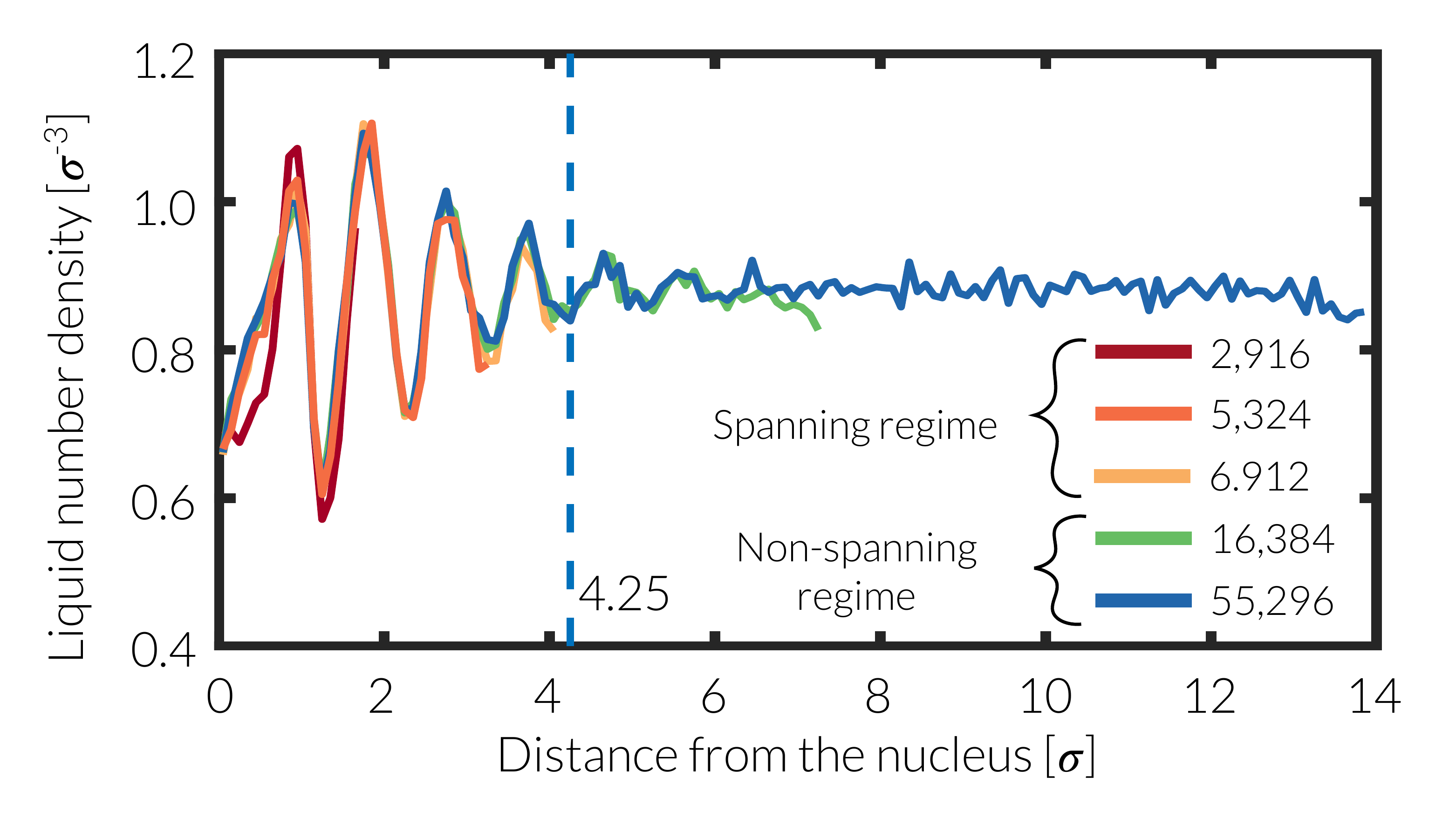}
\vspace{-15pt}
\caption{$\rho_{\bold{u}}(r)$, the inter-image liquid density, as a function of the distance form the critical nucleus for a few representative system sizes.}
\label{Figure 3}
\vspace{-10pt}
\end{figure}

It must, however, be noted that the absence of spanning critical nuclei does not preclude the existence of finite size effects altogether, as the mere proximity of a precritical nucleus to its periodic images can lead to unphysical confinement and  noticeable artifacts. In our earlier work,\cite{HussainJCP2021} we devised a metric for such proximity by analyzing the structural features of the inter-image liquid, as defined in Section~\ref{methods:characterizing proximity}. Fig.~\ref{Figure 3} depicts the inter-image liquid density profiles, $\rho_{\mathbf{u}}(r)$, for non-spanning critical configurations of  several representative systems from both the spanning and the non-spanning regimes. For $r \le 4.25\sigma$,~i.e.,~close to the crystalline nuclei, the liquid is structured as evident from the four peaks of $\rho_{\mathbf{u}}(r)$. Such structuring, however, disappears beyond $r_{c, p} = 4.25\sigma$ wherein $\rho_{\mathbf{u}}(r)$ plateaus, irrespective of  system size. We therefore denote the configurations that lack such a bulk-like plateau region,~i.e.,~those with $|\mathbf{u}| \leq 2r_{c, p}$, as \emph{proximal}.  According to this definition, all spanning configurations are also proximal. Fig.~\ref{Figure 1}b depicts the fraction of proximal critical nuclei as a function of system size. Clearly, all systems in the non-spanning regime harbor critical nuclei that are neither spanning nor proximal while most systems in the spanning regime have a high fraction of proximal critical nuclei. What is intriguing though is that the two largest systems in the spanning regime (i.e.,~$N_p=5,324$ and $N_p=6,912$) have a minuscule ($\leq0.1\%$) fraction of spanning but a considerable fraction of proximal critical nuclei. Nonetheless, they both harbor nucleation at rates that are statistically indistinguishable from those of the larger systems in the non-spanning regime. This is in contrast to heterogeneous ice nucleation in which systems with a large fraction of proximal critical nuclei exhibit significant deviations from the thermodynamic limit, \cite{HussainJCP2021} and suggests that our utilized measure of proximity might not be as strong of an indicator of finite size effects in the LJ system as in the mW system.

\begin{figure*}
\centering
\vspace{-10pt}
\includegraphics[width=0.73\textwidth]{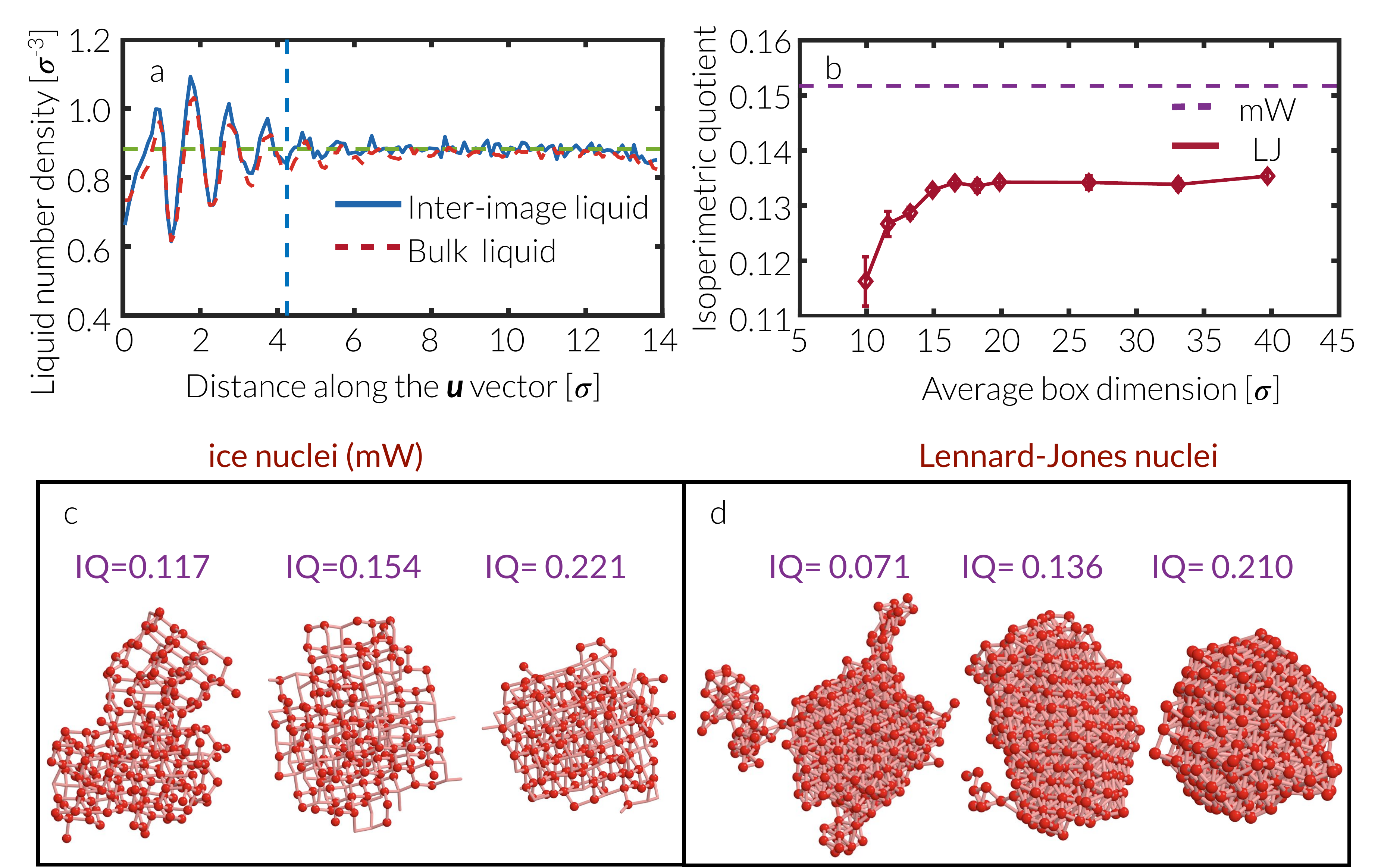}
\caption{(a) The inter-image density profile for the largest system (blue) is very similar to the liquid density profile along randomly chosen vectors of equal lengths connecting pairs of liquid-like particles within the supercooled liquid (red).  (b) Average isoperimetric quotients for critical nuclei within the LJ (diamonds) and mW (dashed line) systems. For mW, the average IQ is computed for critical nuclei collected during homogeneous ice nucleation at  230~K and 1~atm (Ref.~\citenum{HajiAkbariJCP2018}). (c, d) Representative critical nuclei with different IQ values depict the visual distinctions between fragmented (low-IQ ) and compact (high-IQ) crystallites for homogeneous crystal nucleation in (c) mW and (d) LJ systems.}
\label{Figure 4}
\end{figure*}

In order to determine why our proximity measure is such a poor predictor of finite size effects,  we first visually inspect a few representative critical nuclei in the LJ system. As can be seen in Fig.~\ref{Figure 4}d, these nuclei appear to be fragmented,~i.e.,~they are not compact and are instead comprised of elongated appendages and connections between smaller and more compact domains. Such fragmentation is particularly notable for the nucleus labelled as "$\text{IQ}=0.071$`` in Fig.~\ref{Figure 4}d. To rigorously quantify such fragmentation, we use the approach described in Section~\ref{section:methods_IQ} to calculate the average isoperimetric quotient (IQ) of critical nuclei, which is a measure of their sphericity and compactness. A more compact nucleus with a smaller liquid exposed surface area will have a larger IQ value as compared to an elongated and fragmented nucleus. Fig.~\ref{Figure 4}b depicts the average IQ values for critical nuclei of all system sizes considered in this work.
As a point of comparison, we also depict in the dashed line the average IQ value for critical ice nuclei obtained during homogeneous ice nucleation in the mW system~\cite{MolineroJPCB2009} at $T=230~\text{K}$ and $P=1~\text{atm}$.\cite{HajiAkbariJCP2018} This choice allows us to directly compare cluster compactness across different systems that undergo the same mode of nucleation. More specifically, IQ is not a good measure of compactness for  the spherical cap-like nuclei that form during heterogeneous nucleation, which makes any direct comparisons with Ref.~\citenum{HussainJCP2021} impossible. As can be seen in Fig.~\ref{Figure 4}b, IQ values are consistently smaller in the LJ system. In particular, they are almost 11\% smaller for the system sizes  for which $\lessapprox1\%$ of critical nuclei are spanning,~i.e.,~$N_p \geq 4,000$, pointing to less compact and more fragmented nuclei.  In order to visually depict the difference in nucleus fragmentation, several representative critical nuclei with different IQ values are shown in Figs.~\ref{Figure 4}c-d for the mW and LJ systems, respectively. These snapshots make it abundantly clear that nuclei with lower IQ values are indeed less compact and more fragmented. 

How can the compactness of critical nuclei impact the predictiveness of our proximity measure vis-a-vis finite size effects? Intuitively, one would expect the apparent structuring of the inter-image liquid to be less consequential when the nuclei are more fragmented. More precisely,  when a nucleus is  compact, its inter-image vector is  more likely to  connect particles that are located at real two-dimensional (flat or curved) solid-liquid interfaces. Consequently, the extension of the diffuse solid-liquid interface into the inter-image region will induce significant structuring. In the case of more fragmented nuclei, however, the inter-image vector is more likely to connect solid-like particles that sit at their appendages, and the liquid that lies along $\mathbf{u}$ will not be significantly different from the bulk liquid. It must be noted that the liquid density profile along any vector that connects two particle within a liquid will have multiple peaks corresponding to the first few coordination shells of each particle. Therefore, the observed structuring in $\rho_{\mathbf{u}}(r)$ will only be important if $\mathbf{u}$ connects two particles located at two-dimensional solid-liquid interfaces. Due to the fragmented nature of crystalline nuclei in the LJ system, there is a striking similarity between the $\rho_{\mathbf{u}}(r)$  of the $N_p=$55,296 system and the corresponding density profile computed for arbitrary pairs of liquid particles separated by $|\mathbf{u}|_{N_p=55,296}$ as shown in Fig.~\ref{Figure 4}a. (No such similarity is observed for heterogeneous ice nucleation in the mW system as can be seen in SI Fig.~S1.) This suggests that our measure of proximity is not as indicative of finite size artifacts when the corresponding critical nuclei are fragmented.  As such, care must be taken in interpreting the proximity measure by quantifying the extent of fragmentation in nuclei and by comparing $\rho_{\mathbf{u}}(r)$ to a control density profile computed for pairs of liquid-like particles separated by $|\mathbf{u}|$. These observations enable us to  explain the dependence of rate on $N_p$ for the two largest systems in the spanning regime, namely $N_p=$5,324 and 6,912. Both these systems have  spanning fractions of $\le0.1\%$, but vastly different fractions of proximal critical nuclei.

\subsection{The non-spanning regime}

\begin{figure}
\centering
\vspace{-5pt}
\includegraphics[width=0.47\textwidth]{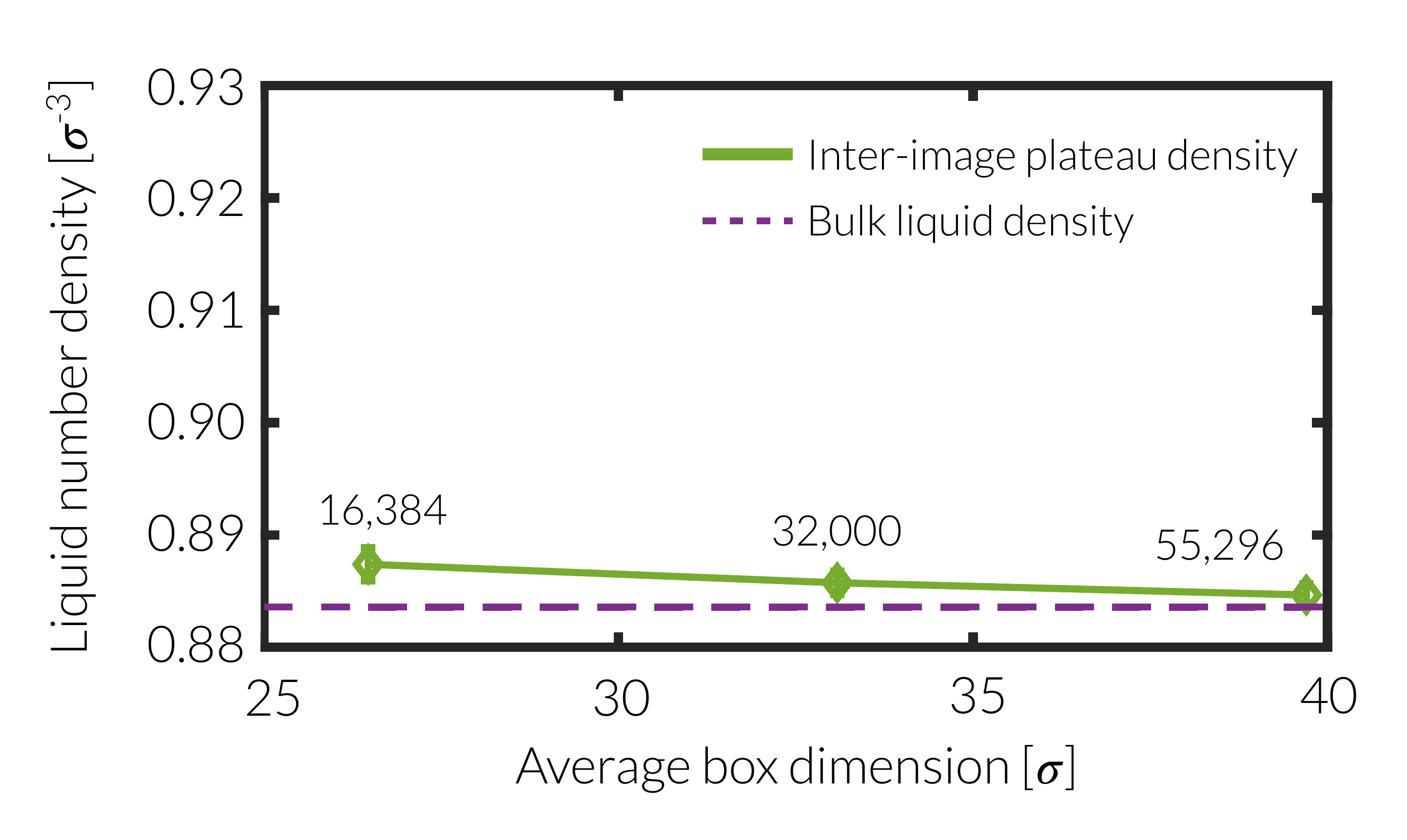}
\vspace{-10pt}
\caption{Average densities of the inter-image liquid in the plateau region of $\rho_{\textbf{u}}(r)$, computed for systems in the non-spanning regime. The $N_p$ value for each systems is shown as a label next to the corresponding data point.}
\label{Figure 5}
\vspace{-15pt}
\end{figure}

\noindent
The three largest systems belong to the non-spanning regime since none of their critical nuclei span the periodic boundary nor are they even proximal according to the  measure described above (Fig.~\ref{Figure 1}b). As discussed in our earlier publication,\cite{HussainJCP2021} however, finite size artifacts might  exist even in the absence of proximal critical nuclei if the properties of the liquid that lies within the plateau portion of the inter-image region deviates significantly from those of the supercooled liquid.  Consistent with our earlier work, we compute a weighted average of the inter-image plateau density for systems in the non-spanning regime using the approach described here and compare them to the bulk liquid density at $T^*=0.5$ and $p^*=0$. For each configuration $x\in\mathcal{Q}$ with an inter-image vector $\mathbf{u}(x)$, we first enumerate $N_p(x)$, the number of particles that lie within a cylinder of radius $r_{c,n}$ and length $|\mathbf{u}(x)|-2r_{c,p}$ along $\mathbf{u}(x)$ and centered midway between the nucleus and its closest periodic image. We then obtain its weighted plateau density as,
\begin{eqnarray}
\rho_p(x) &=& \frac{N_p(x)}{\pi r^2_{c,n} (|\mathbf{u}(x)| - 2r_{c, p})} \\
\rho_p &=& \frac{\sum_{x} N_p(x)\rho_p(x)}{\sum_x N_p(x)}
 \end{eqnarray}
where summation is conducted over all critical configurations.  As depicted in Fig.~\ref{Figure 5}, the $\rho_p$ values within the non-spanning regime are very similar to the mean bulk density of the supercooled liquid, and only deviates from it by less than 1\%. This is in contrast to heterogeneous ice nucleation in the mW system wherein $\rho_p$'s deviate from the supercooled liquid density by at least 5\% in the system that exhibited strong finite size effects despite lack of proximity.\cite{HussainJCP2021} The fact that deviations are much smaller in the LJ system is not surprising considering the observed insensitivity of rate to system size, and the poor predictiveness of proximity as an indicator of finite size artifacts.

The lack of any noticeable size-dependent structural artifacts and the apparent insensitivity of rate to system size in the non-spanning regime, however, do not necessarily guarantee full convergence to the thermodynamic limit. Indeed, a weak-- but meaningful-- dependence of rate on system size might still be possible, primarily due to the power-law dependence of key thermodynamic properties such as chemical potential difference between the two phases and the solid-liquid surface tension on system size.\cite{AguadoJChemPhys2001, OreaJCP2005, MastnyJChemPhys2007, BiscayJCP2009, BurtJPhysChemC2016}  For instance, we observed a linear scaling of $\log_{10}{\mathcal{J}}$ with $1/L$, the characteristic size of the ice nucleating particle, in the case of heterogeneous ice nucleation in the mW system.\cite{HussainJCP2021} In this work, however, we can neither establish nor rule out the existence of such a linear (or power-law) scaling for the following two reasons. First of all, the small number of systems in the non-spanning regime will make any analysis prone to overfitting and far from robust.  Moreover, the apparent insensitivity of rate to $N_p$ suggests that such a scaling will likely be too weak to be reliably detected from our rate estimates at their current levels of uncertainty. Indeed, a linear regression between $1/\langle L\rangle$ and $\log_{10}\mathcal{J}$ for the three non-spanning systems yields a decent correlation ($R^2=0.77$) but with massive uncertainties for $\log_{10}\mathcal{J}_{\infty}$ ($-18.02\pm4.76$). A reliable and robust exploration of this possibility therefore requires conducting a larger number of simulations at a considerably elevated level of accuracy. It must, however, be noted that such an undertaking might be of limited practical significance considering the weak dependence of $\log_{10}\mathcal{J}$ on $N_p$.

\section{Conclusions}
\label{section:conclusions}

\noindent
In this work, we use molecular dynamics and jumpy forward flux sampling to assess the sensitivity of homogeneous crystal nucleation rate in the LJ system on $N_p$, the number of particles.  Similar to our earlier work on heterogeneous ice nucleation,\cite{HussainJCP2021} we find that finite size effects arising from periodic boundary conditions can drastically alter the estimates of the nucleation rate. In the case of homogeneous nucleation in the LJ system and for the range of system sizes considered here, the rate changes by as much as six orders of magnitude.  We identify two distinct regimes for the dependence of rate on system size based on the existence of critical nuclei that span the periodic boundary. In the spanning regime, a nonzero fraction of  such nuclei are spanning. The five smallest systems ($N_p \leq 4,000$), in particular, possess a considerable fraction of spanning critical nuclei and exhibit the strongest dependence of rate on $N_p$. The two largest systems in the spanning regime ($N_p = 5,3244$ and $ N_p = 6,912$) possess only a minuscule fraction of spanning nuclei and do not exhibit strong finite size effects. This is despite the fact that a significant fraction of critical nuclei are proximal in both of them, suggesting that proximity (defined as apparent structuring of the liquid along the inter-image vector) is not a good indicator of finite size artifacts in the LJ system. We attribute this to the fragmented geometry of crystalline nuclei in the LJ system, which makes such apparent structuring almost irrelevant.

In the non-spanning regime, critical nuclei are neither spanning nor proximal, and the nucleation rate is virtually independent of $N_p$. As discussed in our earlier work on heterogeneous ice nucleation,\cite{HussainJCP2021} weak finite size effects might still exist in the non-spanning regime, leading to a linear dependence of $\log\mathcal{J}$ on $1/L$. Establishing such a linear scaling was, however, not possible here and will require conducting a larger number of simulations in the non-spanning regime. Moreover, such simulations need to be more extensive in order to decrease the uncertainties of the computed rates, as the dependence of rate on system size appears to be far weaker. Such a scaling might also be of less practical importance as the rates computed for the few largest system do not appear to deviate significantly from one another, strong indicating convergence to the thermodynamic limit. 

As we mentioned in Section~\ref{section:intro}, an \emph{ad hoc} heuristic that is commonly used in computational studies of nucleation is the "10\% rule``, which posits that finite size effects can be neglected as long as the critical nucleus size does not exceed 10\% of the total number of atoms/molecules in the metastable liquid. Unlike our earlier work on heterogeneous ice nucleation, this heuristic seems to perform reasonably well in the case of homogeneous nucleation in the LJ system as the five largest systems that exhibit rates that vary by less than one third of an order of magnitude, all have critical nuclei with fewer than 10\% of the total number of particles in each system. It is, however, necessary to note that this  might be fortuitous and the 10\% rule might be violated for homogeneous nucleation in systems with long-range correlations, such as tetrahedral liquids and systems with longe-range electrostatic interactions. We therefore think that our  heuristics (based on spanning tendency, proximity and properties of the plateau inter-image region) constitute a more rigorous and systematic framework for detecting  finite size artifacts in computational studies of  crystal nucleation.

\section*{Supplementary Materials}

\noindent
Supplementary material contains Fig.~S1, the counterpart of Fig.~\ref{Figure 4}a  for heterogeneous ice nucleation in the mW system

\section*{ACKNOWLEDGMENTS}
\noindent A.H.-A. gratefully acknowledges the support from the National Science Foundation CAREER Award (Grant No. CBET-1751971). These calculations were performed on the Yale Center for Research Computing. This work used the Extreme Science and Engineering Discovery Environment (XSEDE), which is supported by National Science Foundation Grant No. ACI-1548562.73.

\section*{Data Availability}
\noindent
The underlying data supporting the findings of this study are available from the corresponding author upon reasonable request.

\bibliographystyle{apsrev}
\bibliography{References}

\clearpage

\appendix

\setcounter{figure}{0}
\setcounter{table}{0}
\renewcommand{\thefigure}{S\arabic{figure}}
\renewcommand{\thetable}{S\arabic{table}}

\section*{SUPPLEMENTARY INFORMATION}

\begin{figure}
\centering
\includegraphics[width=.45\textwidth]{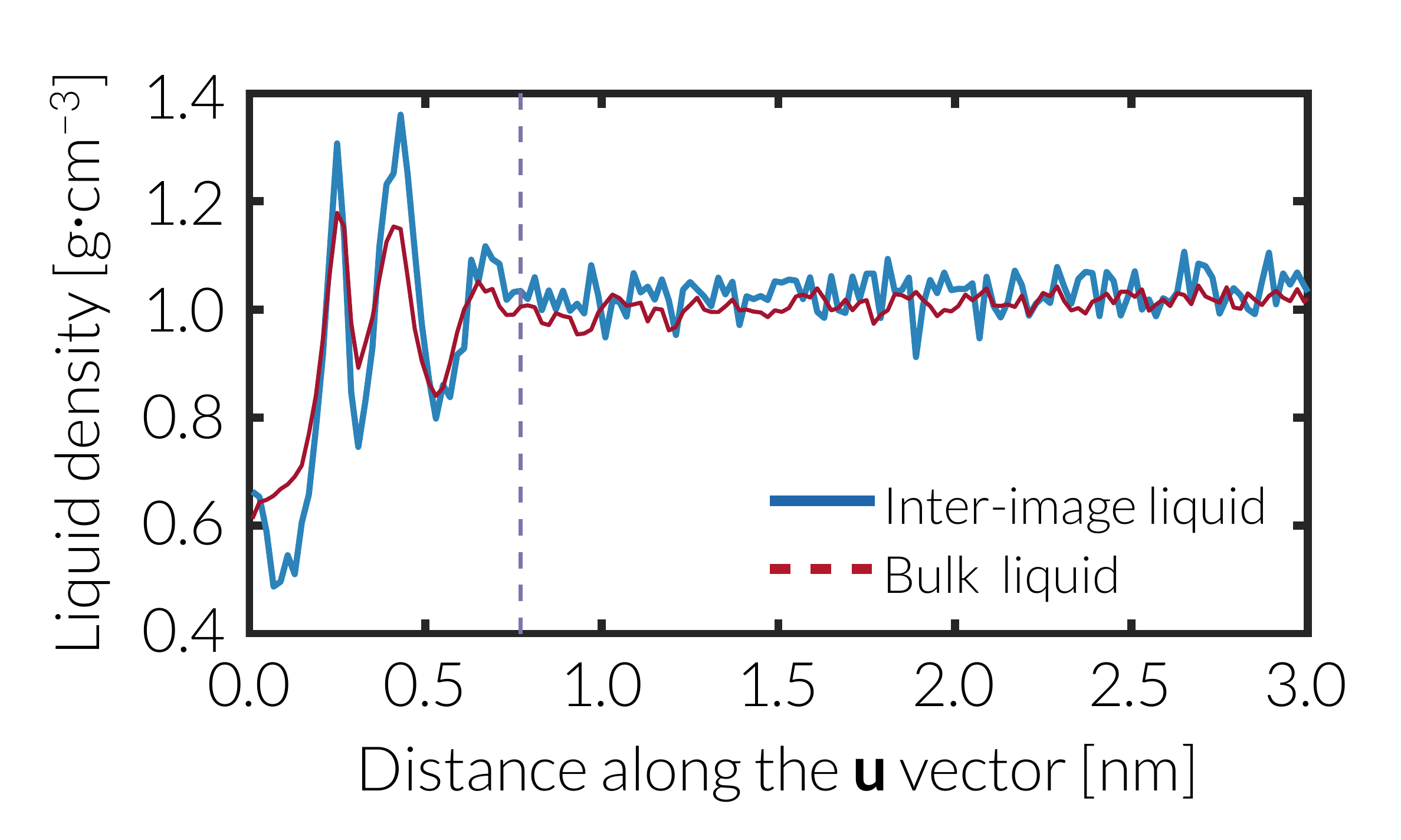}
\caption{In the case of heterogeneous ice nucleation in the mW system, the inter-image density profile for the largest system comprised of 50,176 molecules (blue) is distinctly different from that computed  along  vectors of identical lengths and orientations but connecting pairs of molecules within the supercooled liquid. In order to compute the control curve (red), the molecules connected by each inter-image vector are chosen so that they reside within the same respective liquid layers as those in the nucleating system. }
\end{figure}

\end{document}